\documentstyle[aas2pp4,psfig]{article}

\def\mwd{\hbox{$M_{\rm{}WD}$}}
\def\mpsr{\hbox{$M_{\rm{}PSR}$}}
\def\msun{\hbox{$M_\odot$}}
\def\kms{\hbox{$\rm{}km\,s^{-1}$}}
\let\simgt\gtrsim
\let\simlt\lesssim
\makeatletter
\renewenvironment{deluxetable}[1]{\def\pt@format{\string#1}%
\set@tblnotetext\global\pt@ncol=0\global\pt@column=0\global\pt@page=1%
\def\pt@addcol{\global\advance\pt@ncol by\@ne}}%
{\pt@width\wd\pt@box\box\pt@box\spew@ptblnotes%
\typeout{Page \the\pt@page\space of table \thetable\space has been set to
width \the\pt@width\space with \the\pt@nlines\space lines per page}%
\endcenter\end@float}
\def\startdata{\pt@line=0\pt@calcnlines%
\ifdim\pt@width>\z@\def\@halignto{to \pt@width}\else\def\@halignto{}\fi%
\let\fnum@table=\fnum@ptable\set@mkcaption%
\@float{table}\center\caption{\pt@caption}\leavevmode%
\setbox\pt@box=\pt@tabular{\pt@format}\pt@head}
\def\thebibliography{\subsection*{REFERENCES}
\list{}{\labelwidth3em\leftmargin\labelwidth\labelsep\z@\parsep\z@
\itemsep\z@\itemindent-3em\usecounter{enumi}}
\def\refpar{\relax}
\def\newblock{\hskip .11em plus .33em minus .07em}
\sloppy\clubpenalty4000\widowpenalty4000
\sfcode`\.=1000\relax}
\makeatother

\begin{document}

\title{The masses of the millisecond pulsar J1012+5307 and its
white-dwarf companion\footnote{Based on observations obtained at the
W.~M.~Keck Observatory on Mauna Kea, Hawaii, which is operated jointly
by the California Institute of Technology and the University of
California.}}

\righthead{PSR J1012+5307 and its white-dwarf companion}

\author{M. H. van Kerkwijk\altaffilmark{2,3},
        P. Bergeron\altaffilmark{4}, 
        \and
        S. R. Kulkarni\altaffilmark{2}}

\altaffiltext{2}{Palomar Observatory, California Institute of
                 Technology 105-24, Pasadena, CA 91125, USA; 
                 mhvk, srk@astro.caltech.edu}
\altaffiltext{3}{Hubble Fellow}
\altaffiltext{4}{D\'epartement de Physique, Universit\'e de Montr\'eal,
                 C. P. 6128, Succ. Centre-Ville, Montr\'eal, Qu\'ebec,
                 Canada H3C 3J7; bergeron@astro.umontreal.ca}

\begin{abstract} 

We report on spectroscopy of the white-dwarf companion of the
millisecond radio pulsar PSR~J1012+5307.  We find strong Balmer
absorption lines, as would be expected for a cool DA white dwarf.  The
profiles are much narrower than usual, however, and lines are seen up
to H12, indicating that the companion has a low gravity and hence a
low mass.  This is consistent with the expectation---based on
evolutionary considerations and on the mass function---that it is a
low-mass white dwarf with a helium core.  By comparing the spectra to
model atmospheres, we derive an effective temperature
$T_{\rm{}eff}=8550\pm25\,$K and a surface gravity
$\log{}g=6.75\pm0.07$ (cgs units).  Using the Hamada-Salpeter
mass-radius relation for helium white dwarfs, with an approximate
correction for finite-temperature effects, we infer a mass
$\mwd=0.16\pm0.02\,\msun$.  This is the lowest mass among all
spectroscopically identified white dwarfs.  We determine radial
velocities from our spectra, and find a radial-velocity amplitude of
$280\pm15\,\kms$.  With the pulsar's radial-velocity amplitude, the
mass ratio $\mpsr/\mwd=13.3\pm0.7$.  From all constraints, we find
that with 95\% confidence $1.5<\mpsr/\msun<3.2$.

\end{abstract}


\section{Introduction\label{sec:intro}}

Knowledge of the properties of the white-dwarf companions of radio
pulsars can provide unique constraints on the characteristics and
evolution of these binaries, as well as on those of its constituents.
Of particular interest are mass determinations for millisecond
pulsars.  Evolutionary models indicate that millisecond pulsars have
accreted up to 0.7\,$\msun$ (e.g., Van den Heuvel \& Bitzaraki
\cite{vdheb:95}).  Hence, if the neutron star started with the
``canonical'' 1.4\,$\msun$ (e.g., Timmes, Woosley, \& Weaver
\cite{timmww:96}; for a recent census, Van Kerkwijk, Van
Paradijs, \& Zuiderwijk \cite{vkervpz:95}), it would now be
$\simgt\!2\,\msun$.  If so, this would strongly favor a stiff equation
of state (EOS) at supranuclear density (e.g., Cook, Shapiro, \&
Teukolsky \cite{cookst:94}).  For softer EOS, like the one recently
proposed by Brown \& Bethe (\cite{browb:94}), such a massive neutron
star would collapse into a black hole.  So far, constraints on the
masses could only be set from Shapiro delay of the pulsed signal.  As
this requires (rare) fortuitous alignment, a reasonable constraint is
available for one system only, PSR~B1855+09, for which Kaspi, Taylor,
\& Ryba (\cite{kasptr:94}) found $\mpsr=1.50^{+0.26}_{-0.14}\,\msun$.

The temperature of the white dwarf---combined with the mass---can be
used to determine the age from theoretical cooling tracks.  The
cooling age gives a limit to the time since mass transfer to the
neutron star ceased.  With mass estimates from the mass function and
approximate cooling models, such limits have been used to show that
magnetic fields in neutron stars do not decay on a $\sim\!10^7\,$yr
time scale (Kulkarni \cite{kulk:86}), as had been argued previously
(for a review, Phinney \& Kulkarni \cite{phink:94}).  Recently, Bell
et al.\ (\cite{bell&a:95}) used a lower limit to the cooling age of
the white-dwarf companion of PSR~J0034$-$0534 to infer that the
pulsar's initial spin period was $\simlt\!0.6\,$ms.  If so, this would
also constrain the structure of neutron stars, excluding the stiffer
EOS and thus making it hard to have masses $\simgt\!2\,\msun$ (Cook et
al.\ \cite{cookst:94}; Haensel, Salgado, \& Bonazzola
\cite{haensb:95}).  Without accurate white-dwarf masses and good
cooling models, however, such arguments remain inconclusive.

Masses and temperatures of DA white dwarfs can be determined
accurately from their spectra (e.g., Bergeron, Saffer, \& Liebert
\cite{bergsl:92}), and therefore we have started a spectroscopic
investigation of all bright companions (see Van Kerkwijk
\cite{vker:96}; currently, six companions with $V\simlt23$ are known).
First results can be found in Van Kerkwijk \& Kulkarni
(\cite{vkerk:95}).

Here, we describe spectroscopy of the white-dwarf companion of the
millisecond pulsar PSR~J1012+5307, discovered recently by Nicastro et
al.\ (\cite{nica&a:95}).  The system's orbital period is 14.5\,h, and
the mass function indicates a low companion mass: $\mwd=0.11\,M_\odot$
for $\mpsr=1.4\,M_\odot$ and inclination $i=90^\circ$.  At the
pulsar's position, Nicastro et al.\ found a faint star on the Palomar
sky survey.  Lorimer et al.\ (\cite{lori&a:95}) confirmed the
positional coincidence, and found that the star had magnitudes and
colors consistent with those of a low-mass white dwarf at the distance
implied by the pulsar dispersion measure ($\sim\!0.5\,$kpc).

\begin{figure*}[t]
{\centering\leavevmode\epsfxsize\textwidth\epsfbox{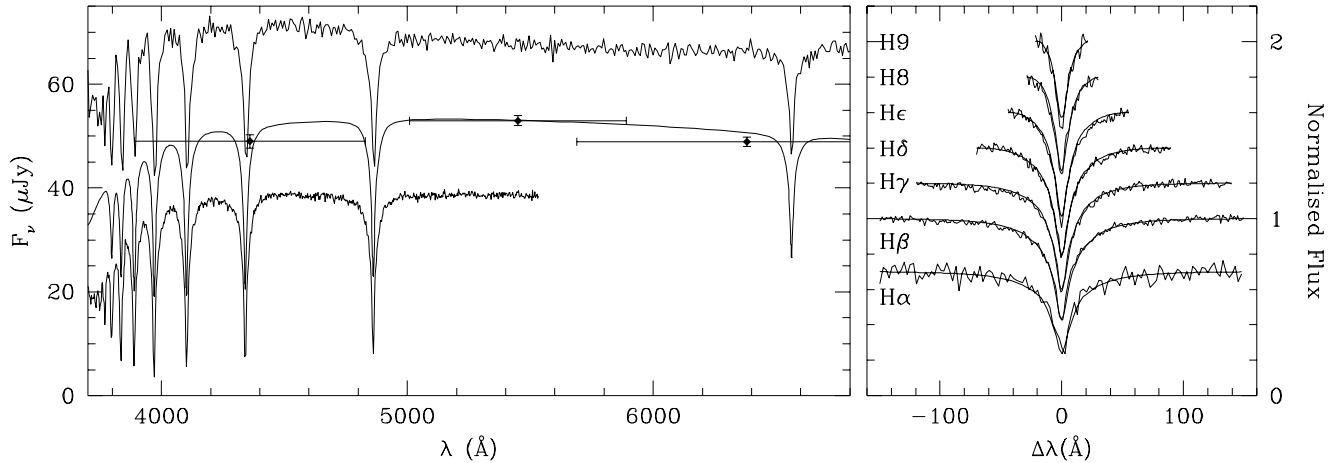}}
\caption[]{The spectrum of the companion of PSR~J1012+5307.  Shown in
the left-hand panel are the spectrum taken with the
300\,line\,mm$^{-1}$ grating (top curve) and the average of the
spectra taken with the 600\,line\,mm$^{-1}$ grating (bottom curve; the
spectra were shifted to zero velocity before averaging).  The two were
normalized to the observed V-band flux ($V=19.58$; Lorimer et al.\
1995), and are shown offset by $+15$ and $-15\,\mu$Jy, respectively.
Also shown are the observed broad-band fluxes and the best-fit pure
Hydrogen model spectrum.  The latter was derived from a fit to the
profiles of H$\beta$ up to H8, and has $T_{\rm{}eff}=8550\pm25\,$K and
$\log{}g=6.75\pm0.07$.  In the right-hand panel, the observed line
profiles, including those of H$\alpha$ and H9, are shown with the
modeled ones superposed.\label{fig:spectrum}\label{fig:proffit}}
\end{figure*}

\section{Observations\label{sec:obs}}

Spectra of the companion of PSR~J1012+5307 ($V=19.5\,$mag) were taken
at the Keck 10\,m telescope with the Low-Resolution Imaging
Spectrometer (LRIS).  On 4 June 1995, the $300\,{\rm{}line\,mm^{-1}}$
grating was used for a classification spectrum, covering the
wavelength range of 3430 to 8460\,\AA\ at $2.5\,{\rm\AA\,pix^{-1}}$.
With a slit width of 1\arcsec, the resolution was $\sim\!12\,$\AA.  The
conditions were photometric, and for calibration an exposure was taken
through a 4\arcsec\ slit.  Directly following the object, the
spectrophotometric standard Feige~34 (sdO, $V=11.23$, Massey et al.\
\cite{mass&a:88}) was observed through both slits, at very similar
airmass.

On 22--25 November 1995 and 21 January 1996, the
$600\,{\rm{}line\,mm^{-1}}$ grating and 0.7\arcsec\ slit were used to
increase the resolution to $\sim\!4$\,\AA.  The wavelength range
covered was 3500--6000\,\AA.  During these runs, the conditions were
not photometric, but Feige~34 was again observed, to obtain relative
spectrophotometry and to verify the stability of the velocity
determinations.

The reduction of all spectra was done using MIDAS\footnote{The Munich
Image Data Analysis System is developed and maintained by the European
Southern Observatory.} and programs running in the MIDAS environment.
The frames were bias-corrected, flat-fielded and sky-subtracted using
standard procedures, and the spectra extracted using optimal
weighting.  Fluxes were determined by first calibrating the wide-slit
exposure relative to that of Feige~34 (correcting for small
differences in airmass using the extinction curve of Beland, Boulade,
\& Davidge \cite{belabd:88}), and then scaling the narrow-slit
exposure such that it had the same continuum flux.

\section{Effective temperature, surface gravity and mass\label{sec:wd}}

The spectra of the companion of PSR~J1012+5307 are shown in
Fig.~\ref{fig:spectrum}.  Strong Balmer lines are seen, from H$\alpha$
up to H12.  In field white dwarfs, the lines are usually broader, and
the higher members of the Balmer series are rarely seen, as the
corresponding excitation levels are quenched by the high pressure.
This indicates that the white dwarf has a low surface gravity and
hence a low mass.

The line profiles could be reproduced well with a white-dwarf model
atmosphere (Fig.~\ref{fig:proffit}), with an effective temperature
$T_{\rm{}eff}=8550\pm25$\,K and a surface gravity
$\log{}g=6.75\pm0.07$ (cgs units).  (For details about the procedure,
see Bergeron et al.\ \cite{bergsl:92}.)  By way of verification, we
also fitted the broad-band fluxes determined by Lorimer et al.\
(\cite{lori&a:95}) to a model atmosphere, with the surface gravity
fixed to $\log{}g=6.75$.  We find $T_{\rm{}eff}=8730\pm110$\,K,
consistent with the spectroscopic determination.

\begin{deluxetable}{ccccc}
\tablewidth{\hsize}
\tablecaption{Mass-radius relation including 
finite-temperature effects\label{tab:mass-radius}}
\tablehead{
\colhead{$M$}&
\colhead{$R_0$}& 
\colhead{$\log{}g_0$}& 
\colhead{$R_{8500}/R_0$}& 
\colhead{$\log{}g$}\\
\colhead{$(M_\odot)$}&
\colhead{$(R_\odot)$}&
\colhead{(cm\,s$^{-2}$)}&
\colhead{}&
\colhead{(cm\,s$^{-2}$)}}
\startdata                             
 0.154& 0.0218& 6.95& 1.351& 6.69\nl   
 0.213& 0.0195& 7.19& 1.236& 7.00\nl   
 0.305& 0.0171& 7.46& 1.111& 7.36\nl   
 0.399& 0.0152& 7.67& 1.071& 7.62\nl   
 0.499& 0.0137& 7.86& 1.048& 7.82\nl   
 0.609& 0.0122& 8.05& 1.032& 8.02\nl   
 0.734& 0.0108& 8.24& 1.022& 8.22\nl   
 0.885& 0.0092& 8.46& 1.010& 8.45\nl   
\enddata
\tablecomments{The first two columns give the zero-temperature
mass-radius relation of Hamada \& Salpeter (1961) for a helium white
dwarf.  The fourth column lists the adopted ratio of the radius at
$T=8500\,$K and the zero-temperature radius, as determined from the
cooling curves of Wood (1995) for carbon white dwarfs with thick
hydrogen envelopes.}
\end{deluxetable}

The mass of the white dwarf can be estimated from the surface gravity
using a mass-radius relation appropriate for a white dwarf with a
helium core (given the low mass, the star will never have burnt
helium).  Unfortunately, helium white dwarfs have not been studied as
well as the usual carbon-oxygen white dwarfs, and only the
zero-temperature mass-radius relation of Hamada \& Salpeter
(\cite{hamas:61}) is available (Table~\ref{tab:mass-radius}).  At
8500\,K, however, a white dwarf with a $\log{}g$ as low as that for
our white dwarf, will not have reached its zero-temperature
configuration.  To estimate the relative ``puffiness'', we have used
the Wood (\cite{wood:95}) models for white dwarfs with pure carbon
cores and thick hydrogen atmospheres (Table~\ref{tab:mass-radius}).
Inter- and extra-polating in the corrected mass-radius relation, our
final estimate for the mass of the white dwarf is
$\log{}M_{\rm{}WD}/M_\odot=-0.78\pm0.06$ (where the quoted uncertainty
combines in quadrature an uncertainty of 0.03 due to the uncertainty
in $\log{}g$, and an estimated 0.05 due to the uncertainty in the
mass-radius relation).  Hence, $M_{\rm{}WD}=0.16\pm0.02\,M_\odot$.

\begin{deluxetable}{rrrrr}
\tablewidth{250pt}
\tablecaption{Radial Velocity Measurements\label{tab:vel}}
\tablehead{
\multicolumn{3}{c}{\dotfill PSR~J1012+5307\dotfill}&
\multicolumn{2}{c}{\dotfill Feige 34\dotfill}\\
\colhead{JD$_{\rm{}bar}$}&
\colhead{$\phi_{\rm{}orb}$\tablenotemark{a}}& 
\colhead{$v$}& 
\colhead{JD$_{\rm{}bar}$}& 
\colhead{$v$}\\
\colhead{$-2450000$}&
\colhead{}&
\colhead{(km\,s$^{-1}$)}&
\colhead{$-2450000$}&
\colhead{(km\,s$^{-1}$)}}
\startdata
$-$127.199& 0.027&$-282\pm27$&  $-$127.184&$ 43\pm10$\nl
    44.040& 0.220&$ -91\pm11$&      44.058&$ 40\pm\phn4$\nl
    44.150& 0.401&$ 194\pm12$&      44.161&$ 20\pm\phn6$\nl
    46.131& 0.678&$  74\pm19$&      46.156&$ 20\pm\phn6$\nl
    46.146& 0.704&$  43\pm10$\nl
    47.097& 0.275&$ -19\pm19$&      47.110&$ 13\pm\phn4$\nl
   103.989& 0.362&$ 138\pm\phn7$&  103.973&$  7\pm\phn4$\nl
   104.014& 0.404&$ 172\pm\phn7$\nl
\enddata
\tablenotetext{a}{Using the ephemeris of Lorimer et al.\
(\cite{lori&a:95}): $T_0={\rm{}JD_{bar}}\,2449220.947499(1)$,
$P_{\rm{}orb}=0.604672713(5)$\,d.}
\end{deluxetable}

\begin{figure}[b]
\plotone{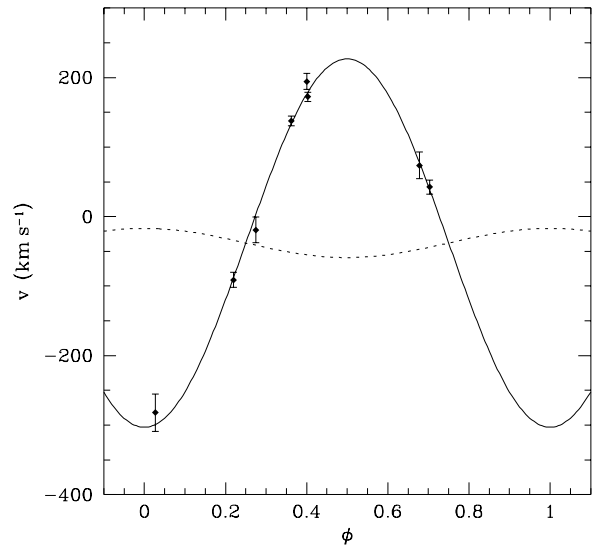}
\caption[]{Radial-velocity measurements for the companion of PSR
J1012+\-5307.  Overdrawn is the best-fit circular orbit, with the time
of ascending node and period fixed to the values determined from the
radio timing solution (Lorimer et al.\ \cite{lori&a:95}).  The
radial-velocity curve for the radio pulsar is shown as well (dotted
curve).\label{fig:velorbit}}
\end{figure}

\section{Radial Velocities and the Mass of the Neutron Star\label{sec:radvel}}

We determined radial velocities for both the companion of
PSR~J1012+5307 and Feige~34 by fitting the H$\beta$, H$\gamma$ and
H$\delta$ lines simultaneously to Lorentzian profiles, with the widths
and velocities of the three lines forced to be the same.  By way
of verification, we also made fits with the velocities left free.  The
results were consistent with each other for all spectra but those
taken with the 300\,line\,mm$^{-1}$ grating.  For Feige~34, we found
that the velocity derived from H$\alpha$ was consistent with the
results from the other spectra, while H$\beta$, H$\gamma$ and
H$\delta$ were off by 60 to 100\,\kms.  Therefore, we used the
velocity from H$\alpha$ alone for J1012+5307 as well.

The velocities were corrected for possible small wavelength
calibration errors using the \ion{O}{1}$\;\lambda5577$ night sky
emission line.  The resulting velocities, corrected to the solar
system barycenter, are listed in Table~\ref{tab:vel}, and shown in
Fig.~\ref{fig:velorbit}.

The radial velocities were fitted to a circular orbit, with the period
and time of ascending node fixed to the radio timing values of Lorimer
et al.\ (\cite{lori&a:95}).  We find a radial-velocity amplitude
$K_{\rm{}WD}=265\pm9\,\kms$ and a systemic velocity
$\gamma=-38\pm6\,\kms$ ($\chi^2_{\rm{}red}=0.97$ for 6 degrees of
freedom).  If we ignore the uncertain velocity from the low-resolution
spectrum, $K_{\rm{}WD}=269\pm11\,\kms$ and $\gamma=-40\pm7\,\kms$
($\chi^2_{\rm{}red}=1.06$ for 5 d.o.f.).

The velocities determined for Feige~34 are, on average, more or less
consistent with the determinations by Greenstein \& Trimble
(\cite{greet:67}, $17\pm13\,\kms$) and Thejll, MacDonald, \& Saffer
(\cite{thejms:91}; $3\pm4\,\kms$). The individual velocities deviate
significantly from the average, however, indicating that some
systematic effects may be present.  To estimate their effect on the
orbital solution, we corrected the velocities for J1012+5307 with
those for Feige~34 (corrected for an assumed velocity for Feige~34 of
$5\,\kms$), and refitted the orbit.  We find
$K_{\rm{}WD}=290\pm9\,\kms$ and $\gamma=-62\pm9\,\kms$, with
$\chi^2_{\rm{}red}=0.73$ (ignoring the low-resolution point:
$297\pm9$, $-66\pm9$ and $0.61$, respectively).  Here, the uncertainty
in $\gamma$ includes a 5\,\kms\ uncertainty in the velocity of
Feige~34.  The $\chi^2_{\rm{}red}$ is lower than for the uncorrected
velocities partly because the errors on the velocities now include the
uncertainty in the velocities derived for Feige~34.  Even if we
neglect the latter, however, $\chi^2_{\rm{}red}$ is still slightly
smaller.  At present, it does not seem possible to decide which method
provides the correct answer, and therefore we will, conservatively,
adopt $K_{\rm{}WD}=280\pm15\,\kms$ and $\gamma=-50\pm15\,\kms$.

\begin{figure}[t]
\plotone{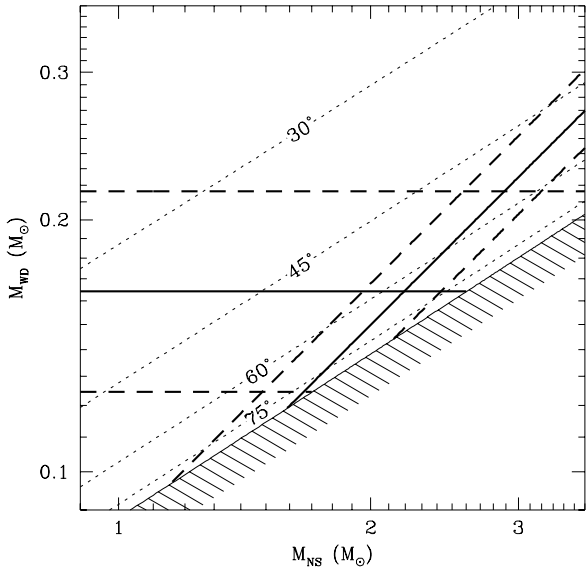}
\caption[]{The masses of PSR~J1012+5307 and its companion.  Shown are
the constraints from the pulsar mass function (thin, drawn curve for
$i=90^\circ$; dotted curves for inclinations as indicated), the ratio
of the radial-velocity amplitudes (thick, slanted curve), and the
white-dwarf surface gravity (thick, horizontal curve).  For the latter
two, the 95\% uncertainty regions are indicated by thick, dashed
curves.\label{fig:masses}}
\end{figure}

{}From the radial-velocity amplitudes of the white dwarf and the pulsar
($20.9774(2)\,\kms$; Lorimer et al.\ \cite{lori&a:95}), it follows
that the mass ratio $\mpsr/\mwd=13.3\pm0.7$.  In
Fig.~\ref{fig:masses}, we show this constraint, as well as the
constraints derived from the pulsar mass function
($5.78361(10)\,10^{-4}\,M_\odot$) and from the white dwarf mass.  From
this figure, it follows that with 95\% confidence the mass of the
neutron star is within $1.5<\mpsr/\msun<3.2$.

\section{Discussion and conclusions\label{sec:disc}}

At present, the neutron-star mass determination is not accurate enough
to constrain the equation of state, or to test the prediction that a
large amount of mass has been accreted.  To improve the accuracy, the
uncertainties in both the mass ratio and the white-dwarf mass need to
be reduced.  The former will be relatively easy (and may well be
interesting on its own; see Fig.~\ref{fig:masses}), and we are
planning more precise radial-velocity measurements for this purpose.

It will be less straightforward to improve the white-dwarf mass
determination.  Although better models are becoming available (Hansen
\& Phinney, in preparation), there are potential problems related to
the uncertainty in the thickness of the hydrogen layer and the unknown
abundance of helium in the atmosphere.  Helium can be dredged up by
convection if the hydrogen layer is thin enough, and, if present,
would mimic an increased surface gravity (Bergeron, Wesemael, \&
Fontaine \cite{bergwf:91}).  Indeed, Reid (\cite{reid:96}) finds that
his mass determinations based on the observations of the Einstein
redshift are systematically lower than the spectroscopic estimates
for temperatures $\simlt\!12000\,$K, which could be attributed to the
presence of helium.  IR observations could potentially resolve this
issue (Bergeron, Saumon, \& Wesemael \cite{bergsw:95}).

If an accurate distance could be determined (via timing parallax or
otherwise), it could be combined with the flux and temperature of the
white dwarf to yield a precise radius.  If different from the radius
predicted from our $\log{}g$ ($0.028\pm0.002\;R_\odot$), one could
assume either that the problem lies with the mass-radius relation (and
not with helium pollution), and infer a mass from the radius and
$\log{}g$, or that there is helium pollution, but that the mass-radius
relation is fine, and use that to derive a mass from the radius.

{}From Fig.~\ref{fig:masses}, one sees that the inclination
$i\simgt\!60^\circ$ for $\mpsr\simlt2\,\msun$.  This raises the
possibility of a model-independent, accurate determination of the
white-dwarf mass and the inclination via Shapiro delay.  With our
precise mass ratio, this would allow one to determine the neutron-star
mass accurately.

Finally, there is a small but measurable amount of light that the
white dwarf will radiate due to reprocessing of the pulsar wind at its
atmosphere.  This may provide yet another constraint on the geometry
of the system.  Fortunately, PSR~J1012+507 is bright in the radio and
the companion relatively bright in the optical, so that all these
observations are possible.

\acknowledgements We thank Brad Hansen, Yanqin Wu, Fernando Camilo and
Andrew Lyne for useful discussions.  M.H.v.K.\ acknowledges a NASA
Hubble Fellowship and S.R.K. grants from NSF, NASA and the Packard
Foundation.

\end{document}